\documentclass[a4paper,11pt]{article}
\pdfoutput=1 % if your are submitting a pdflatex (i.e. if you have
\usepackage{jinstpub} % for details on the use of the package, please
\usepackage{subfig}

\title{\boldmath Characterisation of Large Area THGEMs and Experimental Measurement of the Townsend Coefficients for CF$_4$}

\author[a]{J. Burns}
\author[a,1]{ T. Crane\note{Corresponding author.}}
\author[b]{ A. C. Ezeribe}
\author[a]{ C. Grove}
\author[b]{ W. Lynch}
\author[b]{ A. Scarff}
\author[b]{ N. J. C. Spooner}
 \author[a]{ and C. Steer}

\affiliation[a]{AWE plc, Aldermaston, Reading, Berkshire, RG7 4PR, United Kingdom}
\affiliation[b]{Department of Physics and Astronomy, University of Sheffield, S3 7RH, United Kingdom}

\emailAdd{tom.crane@awe.co.uk}

\abstract{Whilst the performance of small THGEMs is well known, here we consider the challenges in scaling these up to large area charge readouts. We first verify the expected gain of larger THGEMs by reporting experimental Townsend coefficients for a 10~cm diameter THGEM in low-pressure CF$_4$. Large area 50~cm by 50~cm THGEMs were sourced from a commercial PCB supplier and geometrical imperfections were observed which we quantified using an optical camera setup. The large area THGEMs were experimentally characterised at Boulby Underground Laboratory through a series of gain calibrations and alpha spectrum measurements. ANSYS, Magboltz and Garfield++ simulations of the design of a TPC based on the large area THGEMs are presented. We also consider their implications for directional dark matter research and potential applications within nuclear security.}

\keywords{THGEM, micropattern gaseous detectors, time projection chambers, TPC, CF$_4$
\newline
\newline
\newline
\centerline{\copyright~British Crown Owned Copyright 2017/AWE}}

\begin{document}
\maketitle
\flushbottom

\section{Introduction}

Time projection chambers (TPCs) based on micropattern gaseous detectors such as gaseous electron multipliers (GEMs) are a robust and inexpensive radiation detection technology. Applications of GEM-based TPCs are varied, ranging from the detection of weakly interacting massive particles (WIMPs) in the search for dark matter \cite{Phan2017} to medical physics applications such as radiotherapy \cite{Weiden2015}. TPCs are also of interest for nuclear security applications including the detection and imaging of radiological sources and special nuclear material (SNM) \cite{Saenboonruang2014}.

In this paper we present work towards the implementation of a large area thick-GEM (THGEM) based TPC developed jointly by the Atomic Weapons Establishment (AWE) and the University of Sheffield. The dual motivations for this project stem from nuclear security applications of interest to AWE and as a prospective readout technology under consideration by the University of Sheffield for the CYGNUS directional dark matter detector \cite{Spooner2017}.

The TPC's ability to track the three-dimensional trajectories of electrons enables it to reconstruct Compton scattering processes, providing a route for quantitative gamma-ray imaging~\cite{Tanimori2017}. The results of the electron recoil Compton camera from the SMILE collaboration \cite{Smile2007} suggest that gamma-ray sources can be localised with fewer counts than conventional Compton cameras; the resulting operational benefit is to have faster orphan source search times.

The detection of nuclear recoils in the same apparatus offers a flexible instrument for nuclear security applications. This, combined with their directional sensitivity and wide acceptance angle makes them equally attractive for measuring WIMP-induced recoil tracks for directional dark matter detectors such as those currently being constructed by the CYGNUS collaboration \cite{Spooner2017,Miuchi2017}.

THGEMs differ from conventional GEMs due to their fabrication using printed circuit board (PCB) technology \cite{Breskin2009}. As result, they are more robust than regular GEMs which makes them more easily scalable to large-area detectors. Whilst this is typically achieved by tiling smaller GEM boards, the increased tolerance of THGEMs to electrical discharges or sparking makes larger area tiles more practical. The THGEMs described in this work measure 50~cm by 50~cm, considerably larger than THGEMs in earlier reports \cite{Breskin2009,Chechik2006,Cortesi2013} and simpler in design. Boards of this size offer reduced dead-area for large-scale detectors and minimise the impact of electric field perturbations near the edges of the boards.

This report begins with an experimental study of the Townsend coefficients of low pressure CF$_4$ using a 10~cm diameter circular CERN THGEM, followed by an overview of an optical inspection system used to identify THGEM fabrication imperfections. We then present characterisation measurements of our much larger 50~cm~by~50~cm square THGEM boards obtained using an Americium-241 alpha source alongside detector modelling results obtained using ANSYS\cite{2012}, Magboltz\cite{Baraka2015} and Garfield++\cite{Biagi1995}.

\section{Calculation of Townsend Coefficients for CF$_4$}

\subsection{Experimental Setup} 
\label{sec:thgemsetup}

The setup used to study the Townsend coefficients of CF$_4$ employed a small area THGEM above a cathode plane used to establish a drift field. The CERN THGEM electrode was 0.4~mm thick, with a 0.4~mm hole diameter and a pitch of 0.6~mm. Each hole had an etched rim 0.04~mm wide and the distance between the THGEM and the cathode was 2 cm. The parameters of the THGEM are given in Table \ref{THGEMperams}, along with the specifications of the larger THGEMs discussed later in this paper. 

\begin{table}[h]
\centering
\begin{tabular}{|c|cccc|}
\hline
 & Pitch (mm) & Hole Diameter (mm) & Etched Rim (mm) &Thickness (mm) \\ 
\hline
CERN & 0.6 & 0.4 & 0.04 & 0.4 \\
Large Area & 1.2 & 0.4 & 0.15 & 0.47 (0.4 plastic. 0.07 metal) \\
\hline
\end{tabular}
\caption{Parameters of the THGEMs discussed in this work. The size of the etched rim for the large area THGEMs was limited by the manufacturer's tooling.}
\label{THGEMperams}
\end{table}

\begin{figure}[htbp]
\centering
\includegraphics[height=6cm]{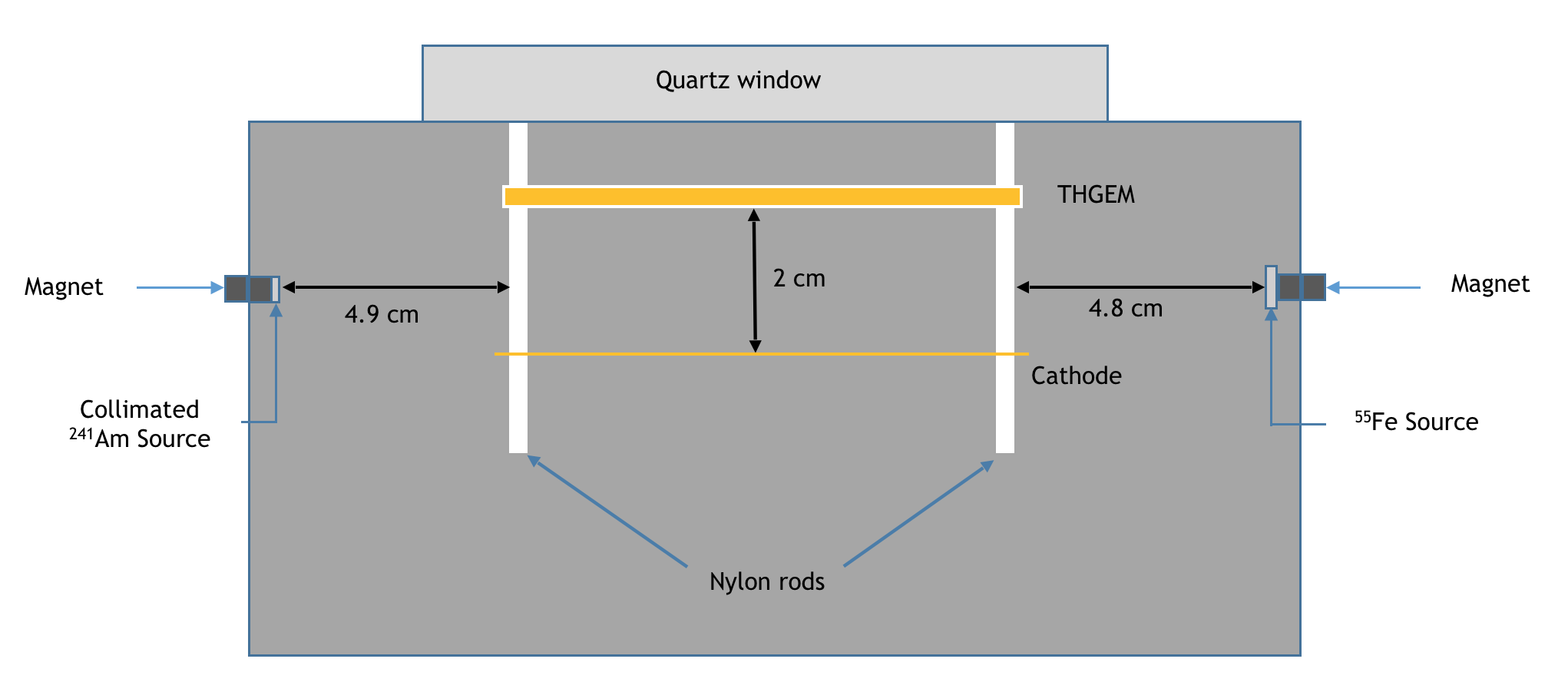}
\caption{\small{Experimental configuration used to measure the Townsend coefficient of CF$_4$. This study utilised a smaller THGEM inside a 10 litre vacuum vessel.}}
\label{fig:thgemgeometry}
\end{figure}

The detector was located inside a 10 litre vacuum vessel filled with low pressure CF$_{4}$ gas. There were two sources in the vessel for calibration and analysis; Americium ($^{241}$Am) and Iron ($^{55}$Fe). Both sources were connected to small magnets which allowed them to be moved into and away from the detection region to give a background result without the need to open the vessel.  A diagram of the geometry inside the vessel is shown in Figure \ref{fig:thgemgeometry}.

The upper plane of the THGEM was held at positive high voltage (HV) and the lower plane was held at ground with the cathode held at negative HV. The signal of the THGEM was read from the upper plane through an Ortec 142 IH preamplifier, before going to an Ortec 572 shaping amplifier and being recorded using an Ortec 926 ADCAM multi-channel buffer (MCB).

\subsection{Gain Calibration}

\begin{figure}[htbp]
\centering
\includegraphics[height=6cm]{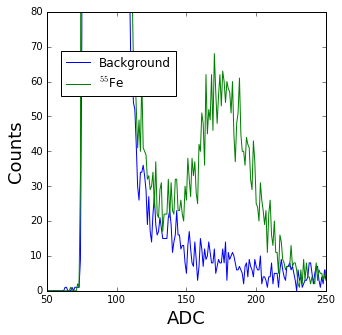}
\caption{\small{Energy spectrum obtained from $^{55}$Fe x-rays (green). A background spectrum (blue) is provided for comparison.}}
\label{fig:fe55spec}
\end{figure}

To calculate the Townsend coefficients of the gas, the gain of the detector is needed for given applied electrostatic fields and pressures. Calibration runs with an $^{55}$Fe x-ray source were used and an energy spectrum was taken using an Ortec-926 MCB. An example energy spectrum, along with a background for comparison, is shown in Figure \ref{fig:fe55spec}. 

\begin{figure}[htbp]
\centering
\includegraphics[height=6cm]{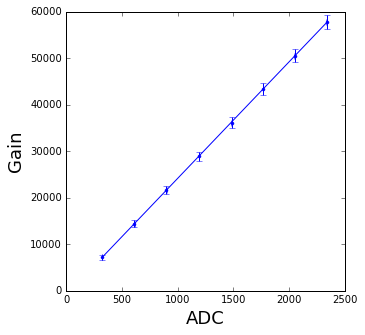}
\caption{\small{Gain against ADC values of the Ortex 926 ADCAM MCB from the calibration with a Tennelec TC 814 pulser.}}
\label{fig:gainvsadc}
\end{figure}

To calibrate the gain a test pulse from a Tennelec TC 814 pulser was sent to the test input of the Ortec preamplifier. Using the known value of the input capacitor of the preamplifier channel (1~pF) the input charge was found, which was then converted into the number of electron-ion pairs using the ionisation energy (W) of CF$_4$  of 34 eV \cite{Reinking1986} and the 5.89 keV energy of the incident x-rays. The gain was found by calculating the ratio of the observed number of electron-ion pairs by the initial number. Following this conversion for all pulse heights leads to the plot of gain against ADC shown in Figure \ref{fig:gainvsadc}. A least squares fit was used on the data of this plot to produce a gain to ADC conversion.

\begin{figure}[htbp]
\centering
\includegraphics[height=6cm]{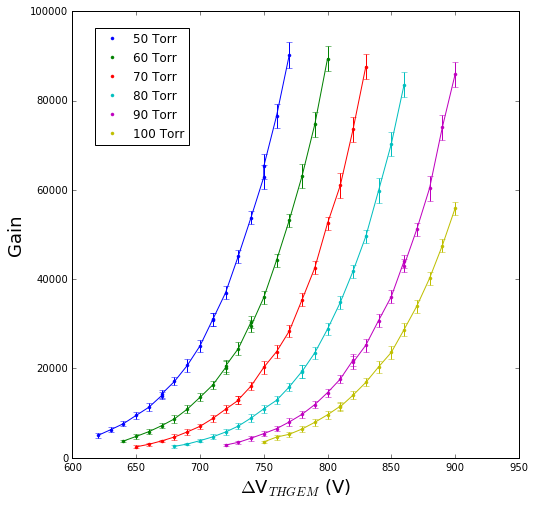}
\caption{\small{THGEM gain against the voltage across the THGEM holes for different pressures of CF$_{4}$ gas.}}
\label{fig:gainvsvoltage}
\end{figure}

After this conversion, a gain curve was made for each different pressure. This is shown in Figure \ref{fig:gainvsvoltage}. The gain can be seen to rise exponentially with increasing voltage, as expected.

\subsection{Calculation of Townsend Coefficients}

Using the values of the gain at differing voltages and pressures of CF$_{4}$, the Townsend coefficients can be calculated from Equation \ref{eq:Towns}.

\begin{equation}
\frac{\alpha}{P} = Ae^{\frac{-B P}{E}}
\label{eq:Towns}
\end{equation}

Where $P$ is the gas pressure, $E$ is the electric field between the plates, $A$ and $B$ are constants and $\alpha = ln\left(Gain\right)/d$ where $d$ is the separation between plates. Rearranging Equation \ref{eq:Towns} leads us to 
Equation \ref{eq:rearrTowns}, from which we can plot $ln\left(ln\left(Gain\right)\right)$ as a function of $1/E$ to find values for the constants $A$ and $B$ from the intercept and gradient respectively.

\begin{equation}
ln\left(ln\left(Gain\right)\right) = ln(APd) - \frac{BP}{E}
\label{eq:rearrTowns}
\end{equation}

\begin{figure}[htbp]
\centering
\includegraphics[height=6cm]{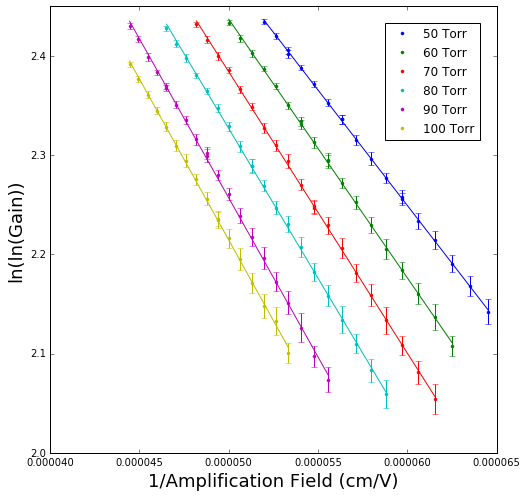}
\caption{\small{ln(ln(gain)) against the reciprocal of the amplification field inside the THGEM holes for different pressures of CF$_{4}$ gas.}}
\label{fig:lnlngain}
\end{figure}

Figure \ref{fig:lnlngain} shows the plot of ln(ln(Gain)) against $1/E$ for each pressure from 50 to 100~Torr and the least squares polynomial fit for each. 

The gradient $m$ and intercept $c$ of the fits can be used along with Equations \ref{eq:A} and \ref{eq:B} to calculate the values for the constants $A$ and $B$ for each pressure of CF$_{4}$. \\

\begin{equation}
A = \frac{e^{c}}{Pd}
\label{eq:A}
\end{equation}

\begin{equation}
B = -\frac{m}{P}
\label{eq:B}
\end{equation}

The values for $A$ and $B$ at different pressures are shown in Table \ref{tab:AandB}. It can be seen that the values of both A and B decrease with increasing pressure, this is the same effect reported in Lightfoot et al. using CS$_{2}$ \cite{Lightfoot2007} and the results are consistent with the only known previous measurement of the Townsend coefficient of CF$_{4}$ shown in Ref. \cite{Arefev1993}.\\

\begin{table}[h]
\centering
 \begin{tabular}{|ccc|}
 \hline
 Pressure (Torr) & A (cm$^{-1}$Torr$^{-1}$) & B (V cm$^{-1}$Torr$^{-1}$) \\ \hline 
 50 & 19.2 $\pm$ 0.5 & 465 $\pm$ 14\\
 60 & 17.6 $\pm$  0.4 & 435 $\pm$ 13\\
 70 & 16.0 $\pm$ 0.4 & 405 $\pm$ 12\\
 80 & 14.5 $\pm$ 0.3 & 377 $\pm$ 11\\
 90 & 13.2 $\pm$ 0.3 & 356 $\pm$ 11\\
 100 & 11.6 $\pm$ 0.2 & 325 $\pm$ 12\\
 \hline
 \end{tabular}
 \caption{Observed values for the gas constants for different pressures of CF$_{4}$.}
 \label{tab:AandB}
 \end{table}
 
\section{THGEM Optical Inspection System}

To better understand and mitigate against sparking we have developed a THGEM inspection system to identify fabrication imperfections which increase susceptibility to sparking. This inspection system has been used to quantify the mechanical precision of boards produced by commercial PCB manufacturers to establish the production yield for comparison to the overall cost of the boards.

\begin{figure}[htbp]
\centering
\subfloat[]
\centering
\includegraphics[width=0.5\textwidth]{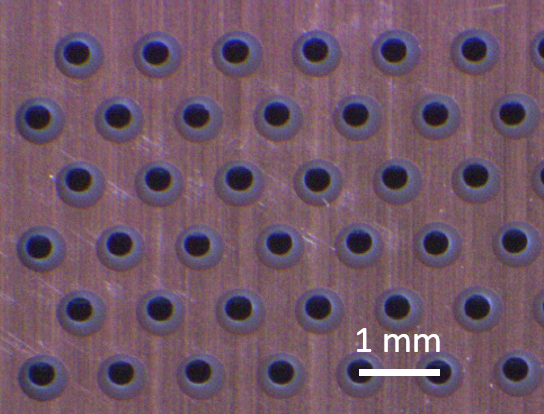}
\subfloat[]
\centering
\includegraphics[width=0.4\textwidth]{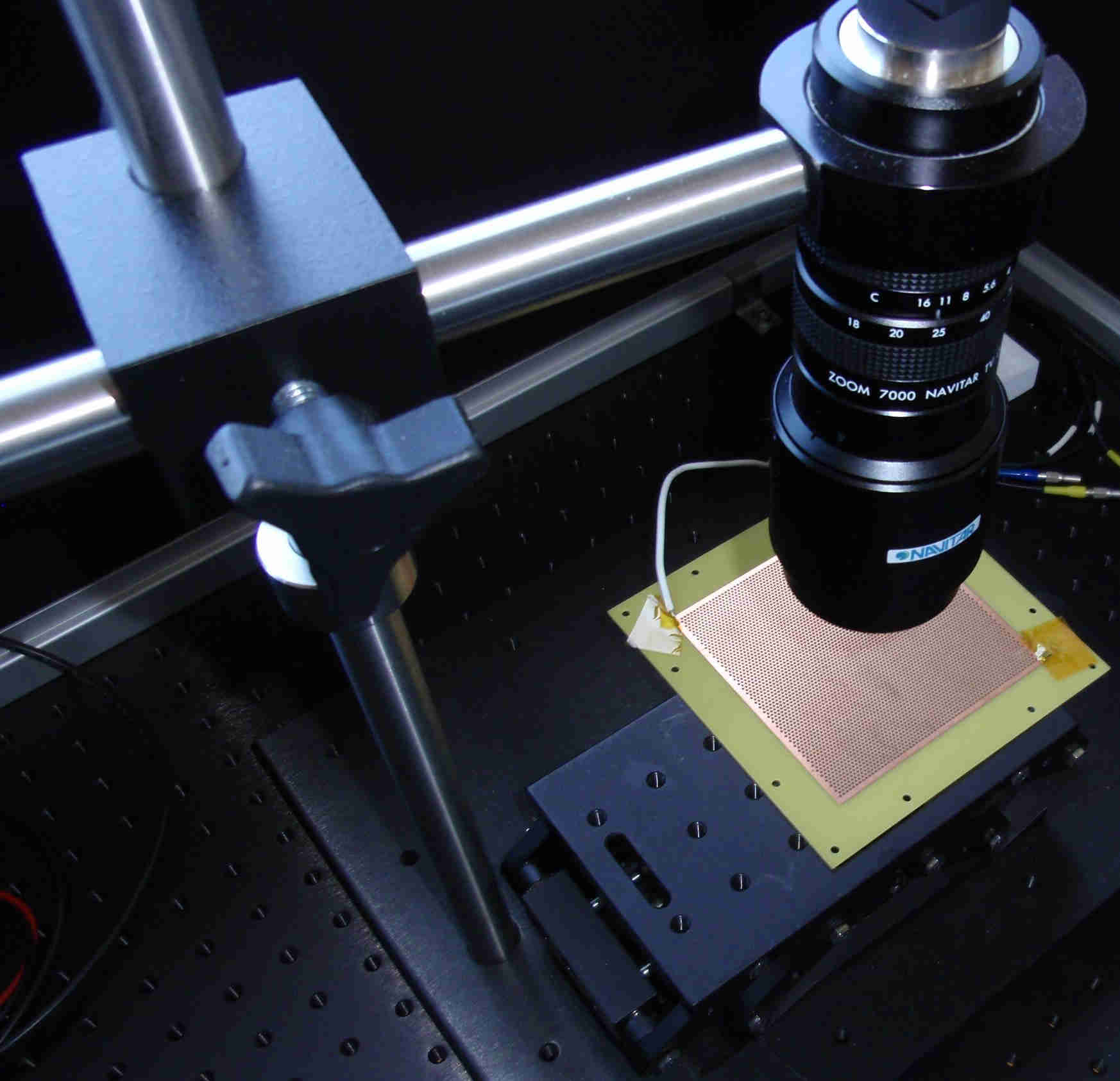}
\caption{Left: Photograph of an area of one of the THGEM boards used in this investigation. Right: Imaging system used on a small-scale prototype for the large area THGEM boards. Coarse movements are made using the camera mount, whilst incremental adjustment is
possible using the translation stage.}
\label{fig:Images}
\end{figure}

Our inspection system comprises a high-resolution CCD camera and image analysis algorithm. First we obtain high resolution images of the THGEM boards, as shown in the left panel of Figure \ref{fig:Images}, which were taken using a camera held by the translatable mounting system shown in the right hand panel.

\begin{figure}[htbp]
 \centering
 \includegraphics[scale=0.25]{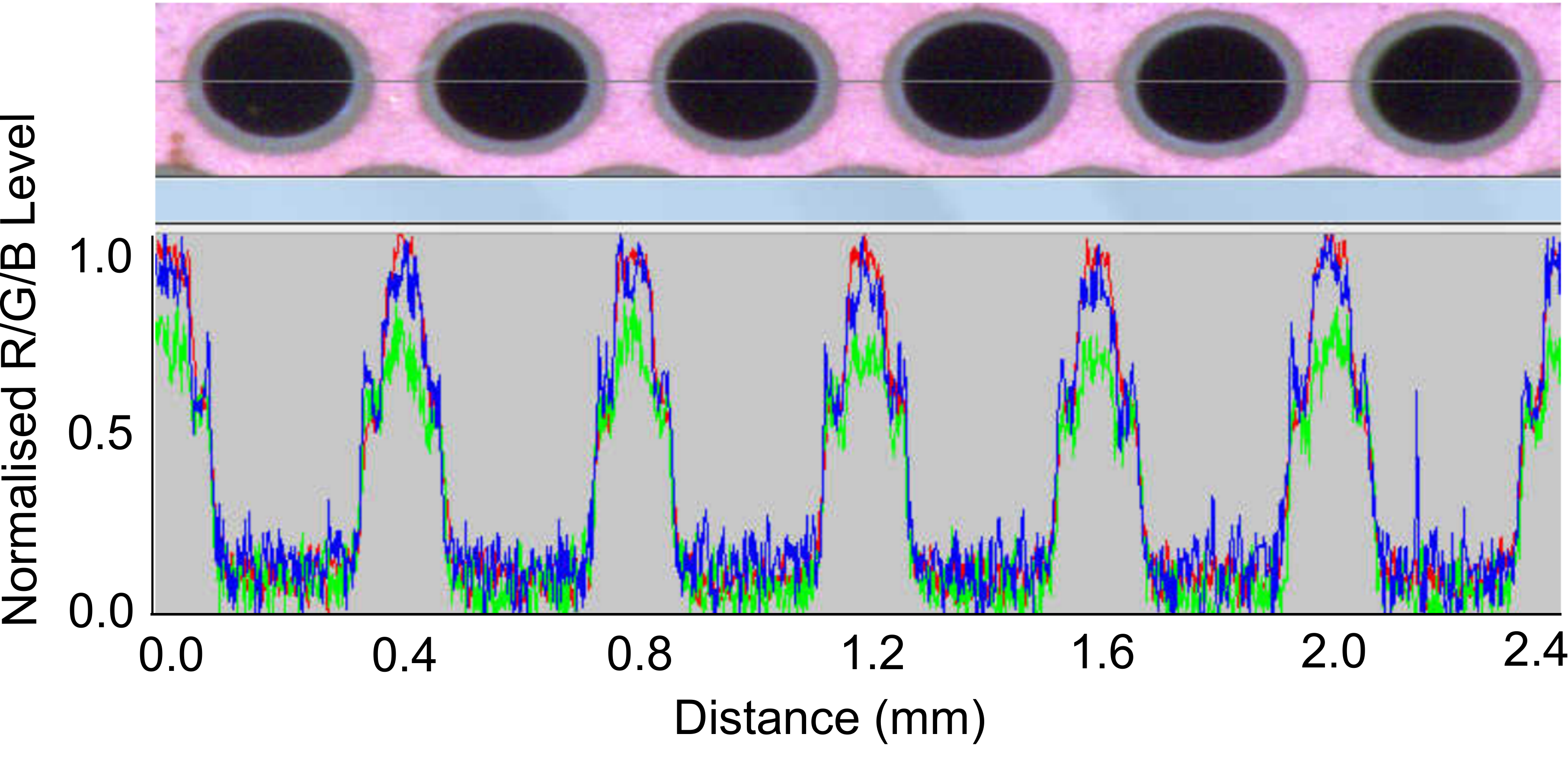}
 \caption{A horizontal line-cut of a THGEM image showing the variation in RGB values across a series of holes. Analysis of the RGB values across two images with different exposure times provides 6 data sets to which edge-detection methods
 can be applied.}
\label{fig:RGB}
\end{figure}

The image analysis code applies an edge detection algorithm that samples red, green and blue (RGB) values from images taken with different exposure times, as shown in Figure~\ref{fig:RGB}. The use of different exposure times permits more accurate discrimination between the copper-coated PCB and the etched areas surrounding each hole.

\begin{figure}[htbp]
 \centering
 \includegraphics[scale=0.25]{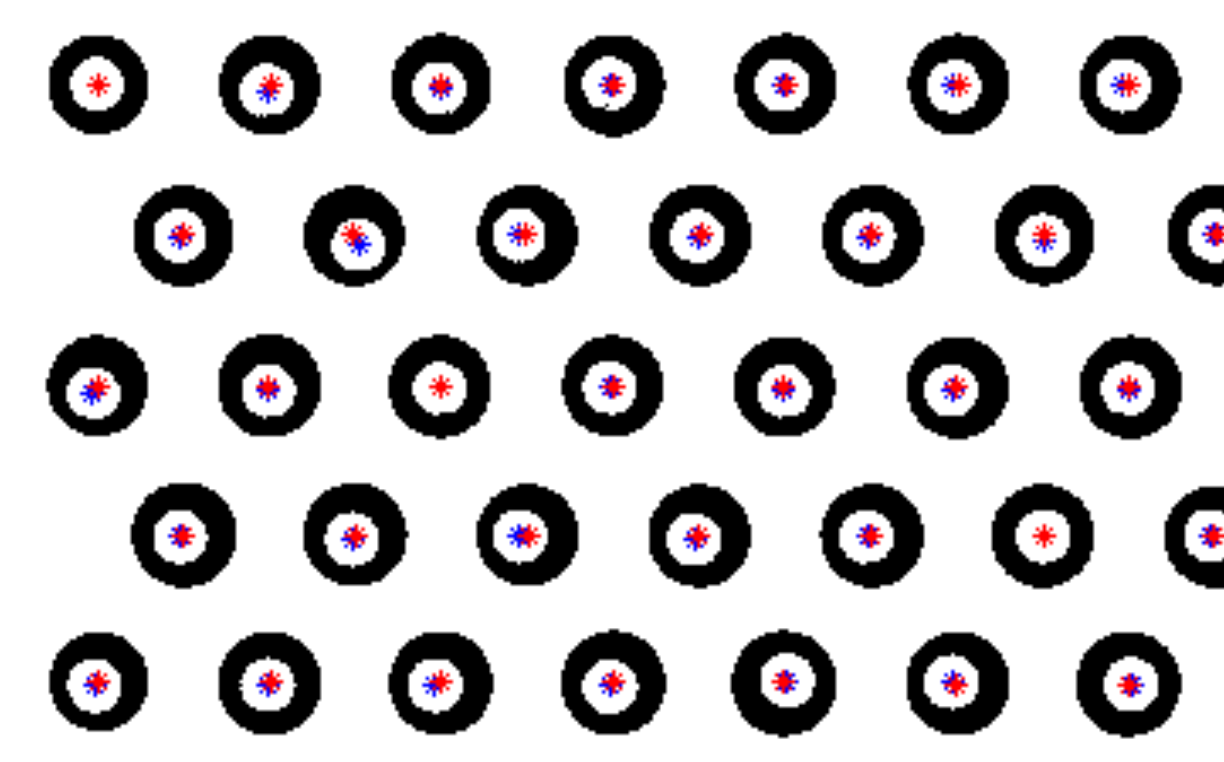}
 \caption{An intermediate result from our image analysis algorithm. The blue and red markers indicate the centres of the drilled holes (white circles) and their corresponding etched rims (black rings) respectively. An offset between the
 two markers indicates that the rim and hole are not perfectly concentric, thereby increasing the susceptibility to sparking at that location.}
\label{fig:CentreDetection}
\end{figure}

The two parameters used to characterise the THGEM fabrication standards were hole displacements and hole eccentricities. Hole displacements indicate that the etched regions around each hole were not perfectly aligned and that the
insulating rims had variable width, as illustrated by Figure \ref{fig:CentreDetection}. Studies of our mechanically drilled boards indicated an average hole-rim displacement of 40~$\mu$m, 27 percent of the total rim width.

Hole eccentricity also reduces the rim width in certain locations and makes sparking more likely. We found an average variation in hole diameter due to eccentricity of 16~$\mu$m. This indicates that hole offsets are the most significant cause of rim width variation. Exploration of other manufacturing techniques for future generations of boards may lead to improved designs with different geometries.

In addition to influencing design choices, our imaging technique also provides an effective method of quality assurance. If any areas of the THGEM boards contain significant deviations in rim width they can be covered with insulating varnish  or polyimide tapes, thus eliminating the risk of sparking at the cost of a small reduction in active area.

\section{Large Area THGEM Experimental Characterisation}

\subsection{Experimental Setup}

\begin{figure}[htbp]
 \centering
 \includegraphics[scale=1.0]{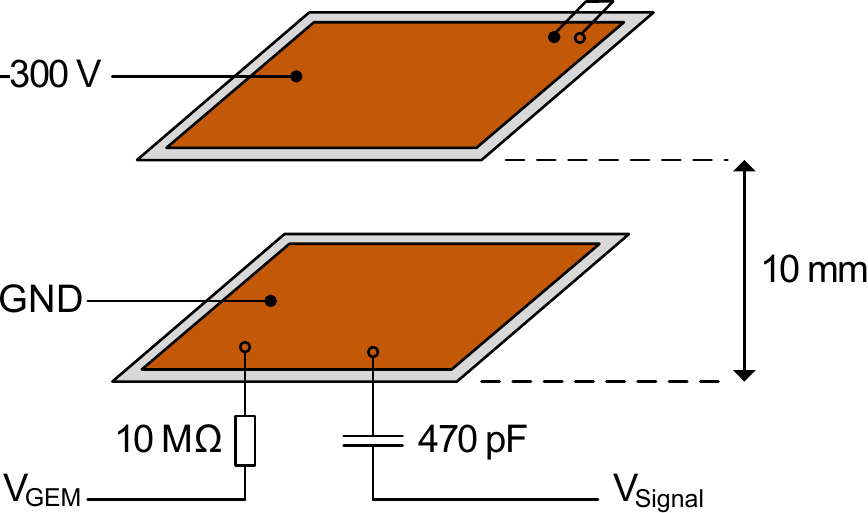}
 \caption{Schematic of the experimental configuration. The cathode (upper board) was held at a fixed potential of -300~V relative to the nearside of the second THGEM board. The far side of the second THGEM board was connected to a
 variable HV supply via a 10 M$\Omega$ resistor to reduce the risk of sparking. The signal output was decoupled from the high voltage board using a 470 pF capacitor.}
\label{fig:Setup}
\end{figure}

After the gain study was completed using small area THGEMs and optical characterisation work performed on a number of different designs, we moved to significantly larger readouts fabricated by a commercial printed circuit board (PCB) manufacturer (QuickCircuits Ltd, UK) with an area of approximately 50~cm by 50~cm.  Preliminary characterisation measurements were performed with an $^{241}$Am source. The THGEM boards were held with a fixed separation of 10~mm, as depicted in Figure \ref{fig:Setup}. The top board functioned as a cathode to establish a drift region whilst the other board served to produce gain from incident electrons and provided an output signal via a capacitor.

\begin{figure}[htbp]
 \centering
 \includegraphics[height=5cm]{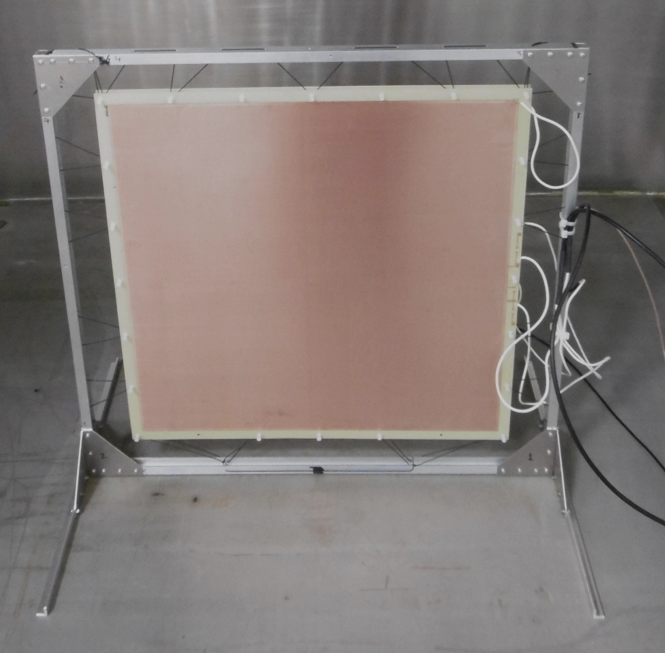}
 \caption{Photograph showing the two THGEM boards supported by an aluminium frame using non-conductive cord. The apparatus is shown within the gas-tight vessel.}
\label{fig:GEM_Photo}
\end{figure}

\begin{figure}[h]
\centering
\includegraphics[height=6cm]{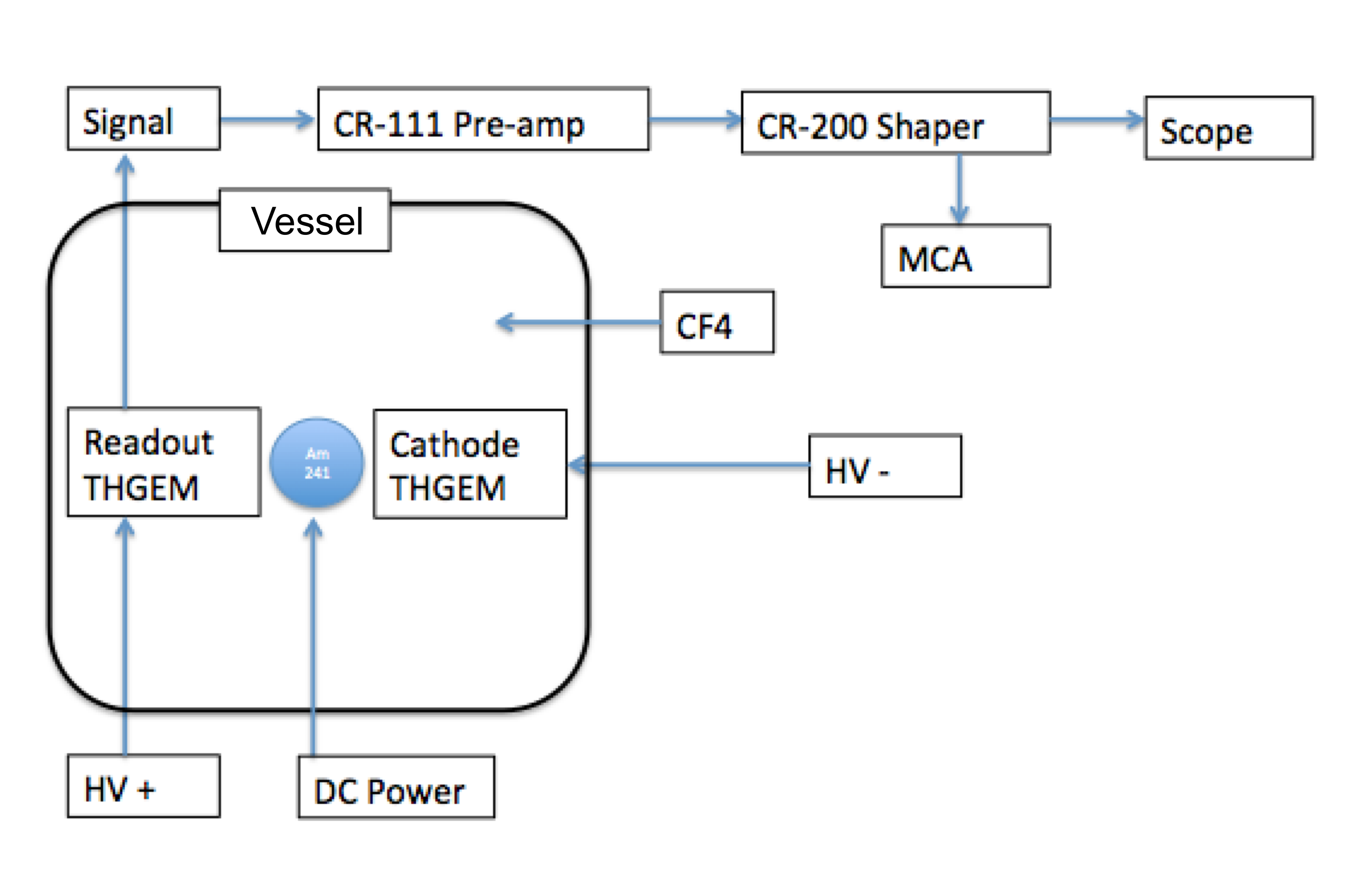}
\caption{A schematic showing the overall experimental configuration and data acquisition chain.}
\label{fig:exp_sch}
\end{figure}

The assembly shown in Figure \ref{fig:GEM_Photo} was placed within the vacuum vessel and a 1~kBq $^{241}$Am sample was placed directly beneath the boards so that as many of the alpha tracks were contained within the drift region as possible. The source holder included a remotely operated shutter and a foil aperture to collimate the alpha emission. The vessel was evacuated and purged with pure CF$_4$ before being pressurised to 30, 50 or 100 Torr and sealed. Figure \ref{fig:exp_sch} shows a schematic of the apparatus used to test the THGEM. The processing electronics consisted of a Cremat CR-111 preamplifier and a CR-200 4$\mu$s shaper.

\subsection{Method}

A bias of -300~V was applied to the cathode whilst the voltage of the second THGEM board, V$_{GEM}$, was increased until regularly occurring pulses from alpha particles were observed on an oscilloscope. An alpha spectrum was then recorded using the multi-channel analyser (MCA), and further measurements were made as the applied voltage was incremented. Cremat amplifiers were calibrated using a test voltage pulse provided by a Tennelec 814 which was passed through a 1~pF capacitor at the pre-amplifier's test input. 

\begin{figure}[h]
\centering
\includegraphics[height=6cm]{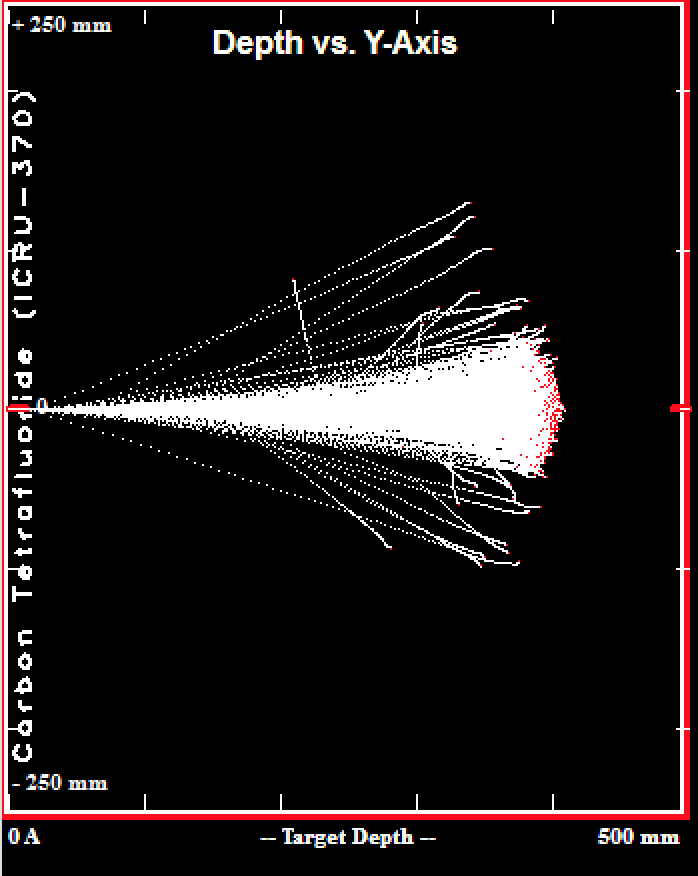}
\includegraphics[height=6cm]{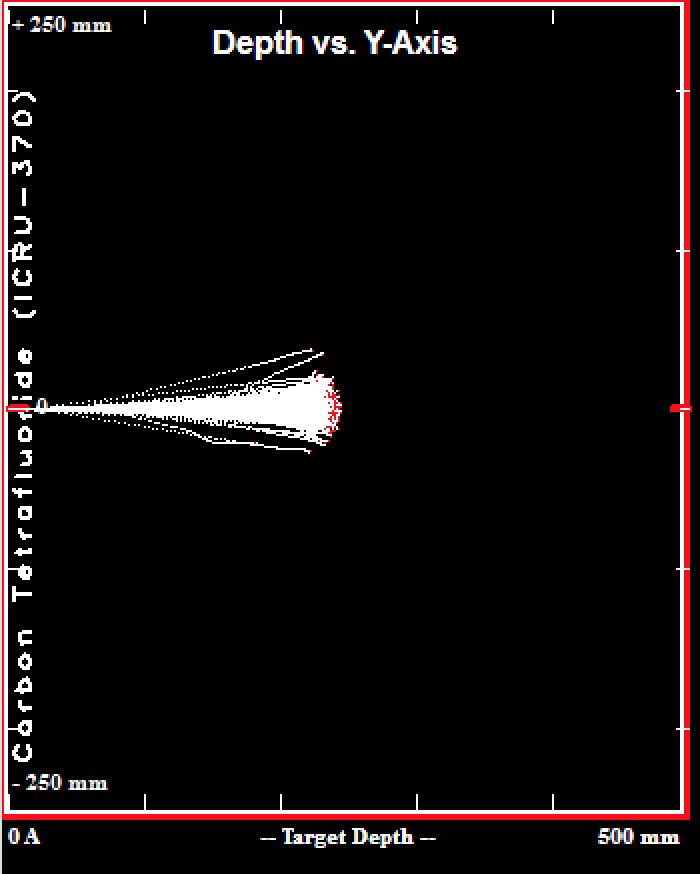}
\caption{SRIM alpha ranges and radial spreads for 30 (left) and 50~Torr (right) CF$_4$.}
\label{SRIM:Ranges}
\end{figure}

The voltage for which the number of electrons was calculated appears as a peak on the MCA. To relate the MCA channel to gain, the amount of energy deposited by the $^{241}$Am alphas into the drift region was estimated using the ion transport software package SRIM \cite{Zeigler2010}. The range of 5.49~MeV alpha particles in 30 and 50~Torr of CF$_4$ was calculated to be 39.3~cm and 25.6~cm respectively. The 50~cm$^2$ THGEM area was sufficient to contain the full alpha range but the 1~cm region between the cathode and readout was not sufficient to contain the radial spread in the alpha tracks which was calculated by SRIM to be 1.0~cm (0.84~cm straggle) and 0.62~cm (0.49~cm straggle) at 30 and 50~Torr respectively. Figure \ref{SRIM:Ranges} shows the alpha range and spread for an incidence angle of zero (simulating collimation) that were used for these calculations at 30 (left) and 50~Torr (right).

To estimate the alpha energy deposited within the drift region SRIM \cite{Zeigler2010} was used to produce a list of alpha energy per position in 3D for 1000 collimated events. The positions were taken at energy steps of 100~keV. The 1~cm gap between the THGEM boards was defined as the z position and all alpha energy deposited outside of this gap  (-0.5 > z > 0.5~cm) was removed from the list. The mean alpha energy for the 1000 events deposited within the 1~cm gap was then calculated. The average fractions of alpha energy deposited within the drift region at 30 and 50~Torr were calculated as 0.89 and 0.95 respectively.   

The SRIM simulations assumed that the $^{241}$Am source was perfectly collimated with a zero angle of incidence. The degree of radial spread due to imperfect collimation will therefore mean a greater loss in alpha energy from the drift region, such that the gain measurements produced from this calibration provide a lower limit of the gain achieved.

The calculated energy fractions were then used to calculate the number of electron-ion pairs produced within the drift region. This information was used to determine the gas gain given by \ref{eq:gain}:

\begin{equation}
gain = {\frac{VCW} {eE_\alpha}}
\label{eq:gain}
\end{equation}

With the capacitance (C), alpha energy ($E_\alpha$) and ionisation energy (W) known, voltage (V) is the only free parameter so entering in the other values gives a gain of 39.03~V.

\begin{figure}[h]
\centering
\includegraphics[height=6cm]{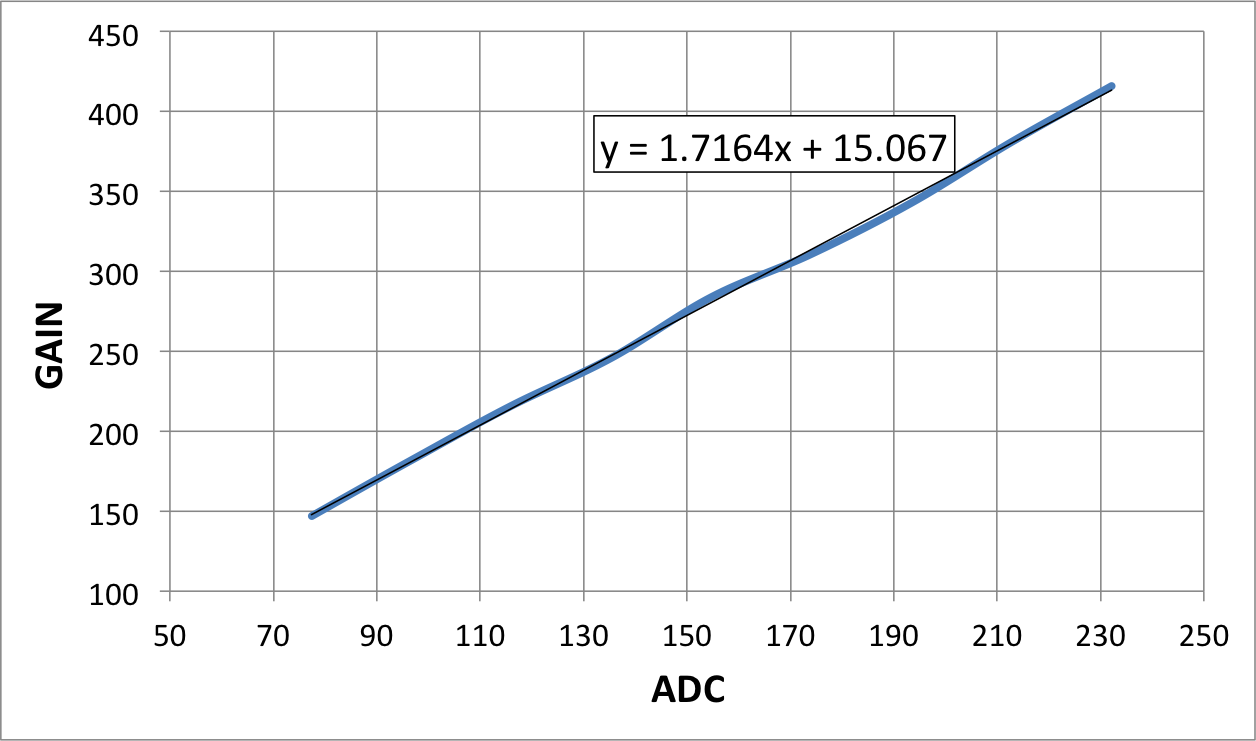}
\caption{Conversion graph for ADC to gain conversion}
\label{fig:conv_graph}
\end{figure}

With this direct relation between gain and voltage a test pulse will correspond to both an MCA channel peak and a gain value. The gain can be plotted against MCA channel and can then be worked out for any channel number. A plot of gain vs ADC (the Analogue-to-Digital Conversion puts the analogue voltage signal into a digital MCA channel) that was derived whilst calibrating this experiment is shown in Figure \ref{fig:conv_graph}. 

It can be seen from Figure \ref{fig:conv_graph} that the relation is linear and a line of best fit produces an equation that can be used to convert the peak ADC value to a gain value. 

\begin{figure}[h]
\centering
\includegraphics[height=5cm]{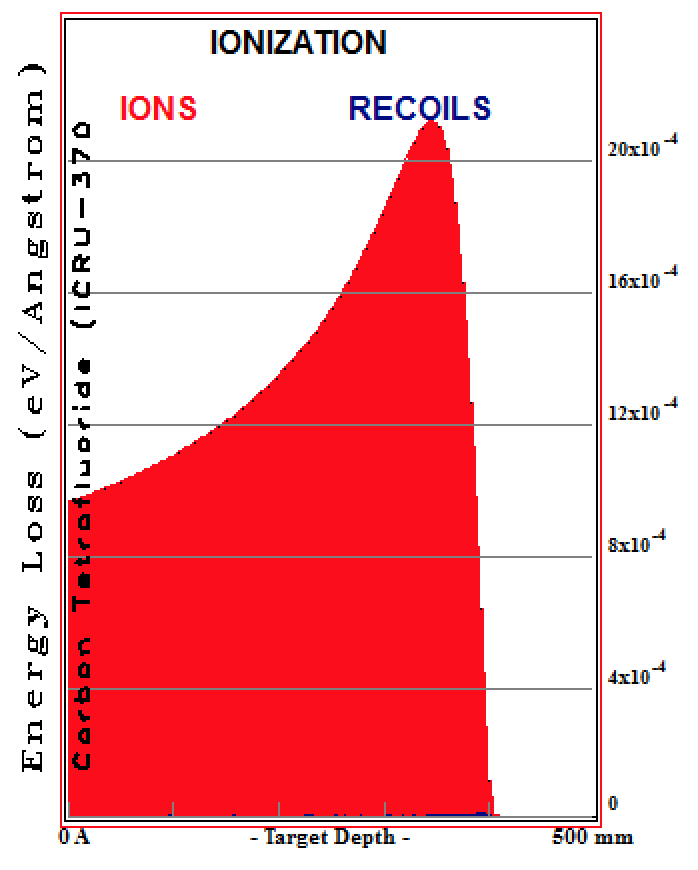}
\includegraphics[height=5cm]{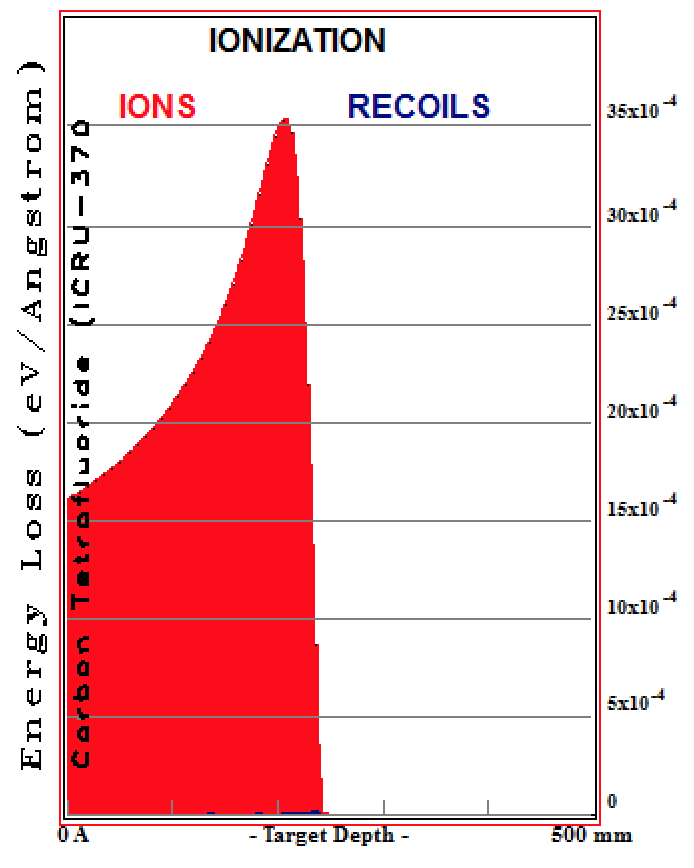}
\caption{Alpha energy distributions calculated using SRIM for CF$_4$ at 30 (left) and 50~Torr (right). Ions produced are shown in red and recoil events are shown in blue (barely visible).}
\label{SRIM:Ionize}
\end{figure}

A further assumption made to calibrate using alphas was that the quenching factor for 5.49~MeV alpha particles in 30 and 50 Torr CF$_4$ is unity. This assumption was supported by a SRIM \cite{Zeigler2010} calculation of the amount of the alpha's initial energy that produces ionisation. This was found to be 99.94\% and 99.74\% for 30 and 50 torr respectively. This is shown graphically in Figure \ref{SRIM:Ionize} where the red region represents the alpha's energy that produces ions and blue (too small to see clearly) is the alpha's energy that causes recoils in the gas.

\subsection{Results}

\begin{figure}[h]
\centering
\includegraphics[height=7cm]{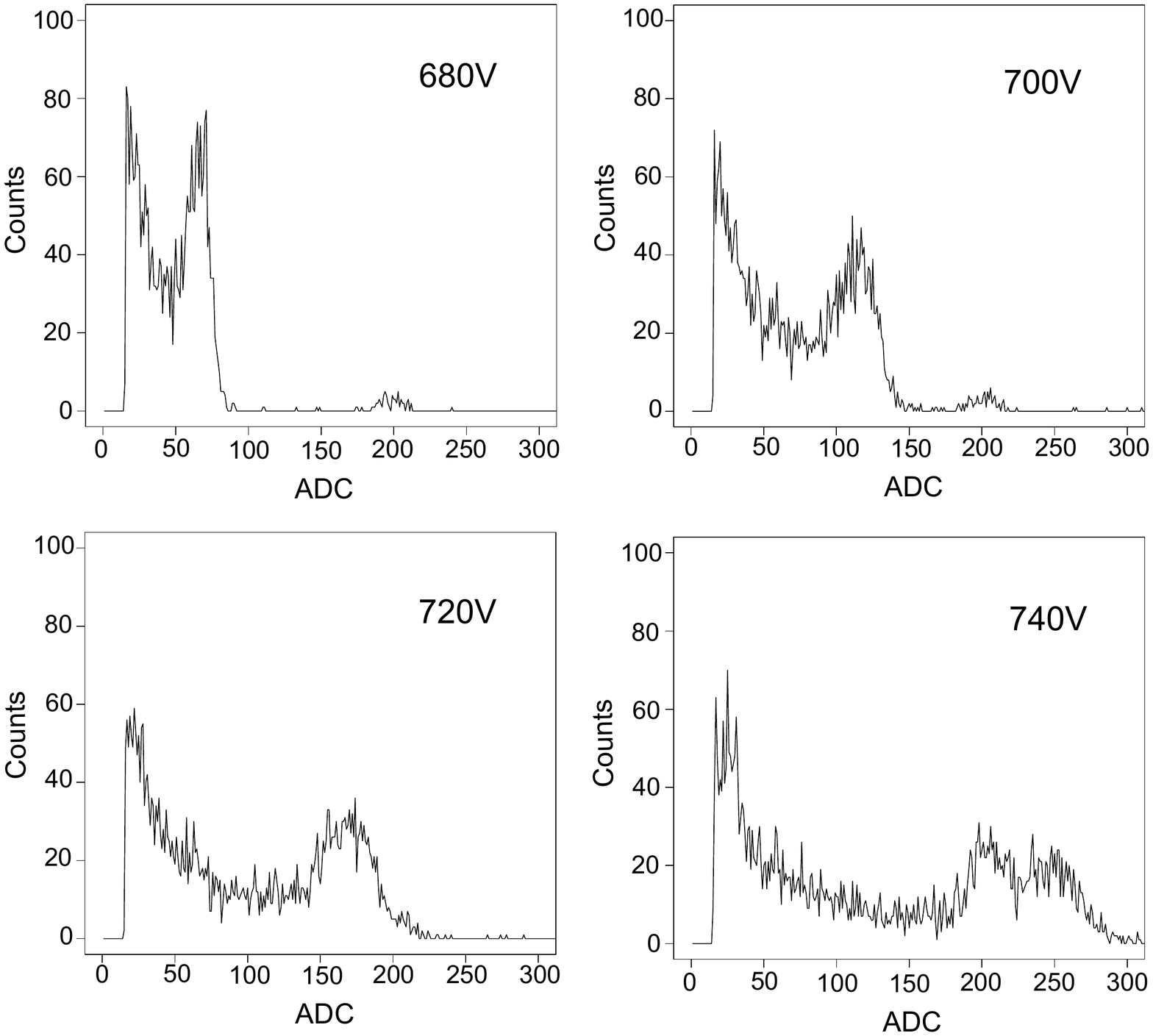}
\caption{Alpha spectra with source on between 680 and 740~V.}
\label{fig:alpha_spectra}
\end{figure}

Figure \ref{fig:alpha_spectra} shows alpha spectra for 50~Torr CF$_4$. The figure shows measurements at 680~V and then shows spectra at steady voltage increases of 20~V up to a maximum of 740~V at which point sparking started to occur at a greater rate which can be seen as a peak in both the source on and off 740~V spectra of Figure \ref{fig:alpha_spectra}.

Further spectra were taken at 30~Torr CF$_4$, then the calibration was used to convert ADC to gain giving the following lower limit gain results shown in Table \ref{THGEMgain} for both 30 and 50~Torr CF$_4$.
\newline

\begin{table}[h]
\centering
\begin{tabular}{|c|cc|}
\hline
& Voltage (V) & Gain\\ 
\hline
30 Torr & 580 & 34\\
& 600 & 51\\
& 620 & 85\\
\hline
50 Torr & 620 & 9\\
& 660 & 24\\
& 700 & 58\\
& 740 & 130\\
\hline
\end{tabular}
\caption{Lower gain limit for the 50~cm~x~50~cm THGEMs.}
\label{THGEMgain}
\end{table}

\section{THGEM and TPC Simulations}

\subsection{ANSYS Modelling}

\begin{figure}[htbp]
\centering
\includegraphics[height=5cm]{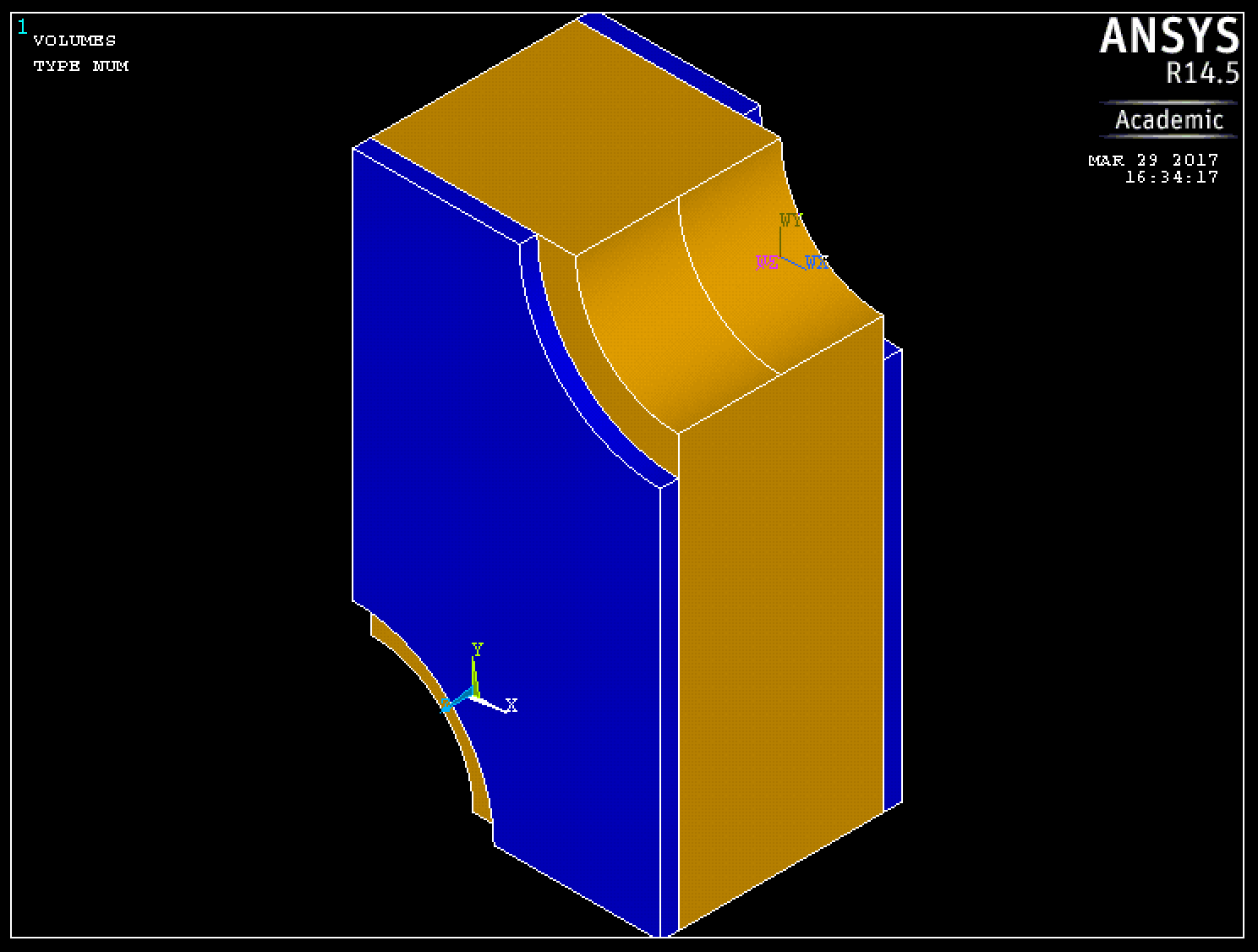}
\caption{Portion of the large area THGEM modelled in Ansys}
\label{fig:Ansys_thGEM}
\end{figure}

Electric field profiles generated by the large area THGEM were modelled using Ansys \cite{2012} by simulating a unit cell and exploiting the periodicity and symmetry of the THGEM design, as shown in Figure \ref{fig:Ansys_thGEM}. The geometry and electric fields generated by Ansys were then exported into Garfield++ \cite{Baraka2015} to perform electron transport simulations.

A gas volume was created as part of the model and relevant voltages were set throughout the volume to create the field. A potential of -300~V was set on the area of the gas that would be touching the cathode. The readout side of the THGEM was set at 650~V and the opposite side was grounded.

\subsection{Garfield++ Gain Simulations}

\begin{figure}[htbp]
\centering
\includegraphics[height=6cm]{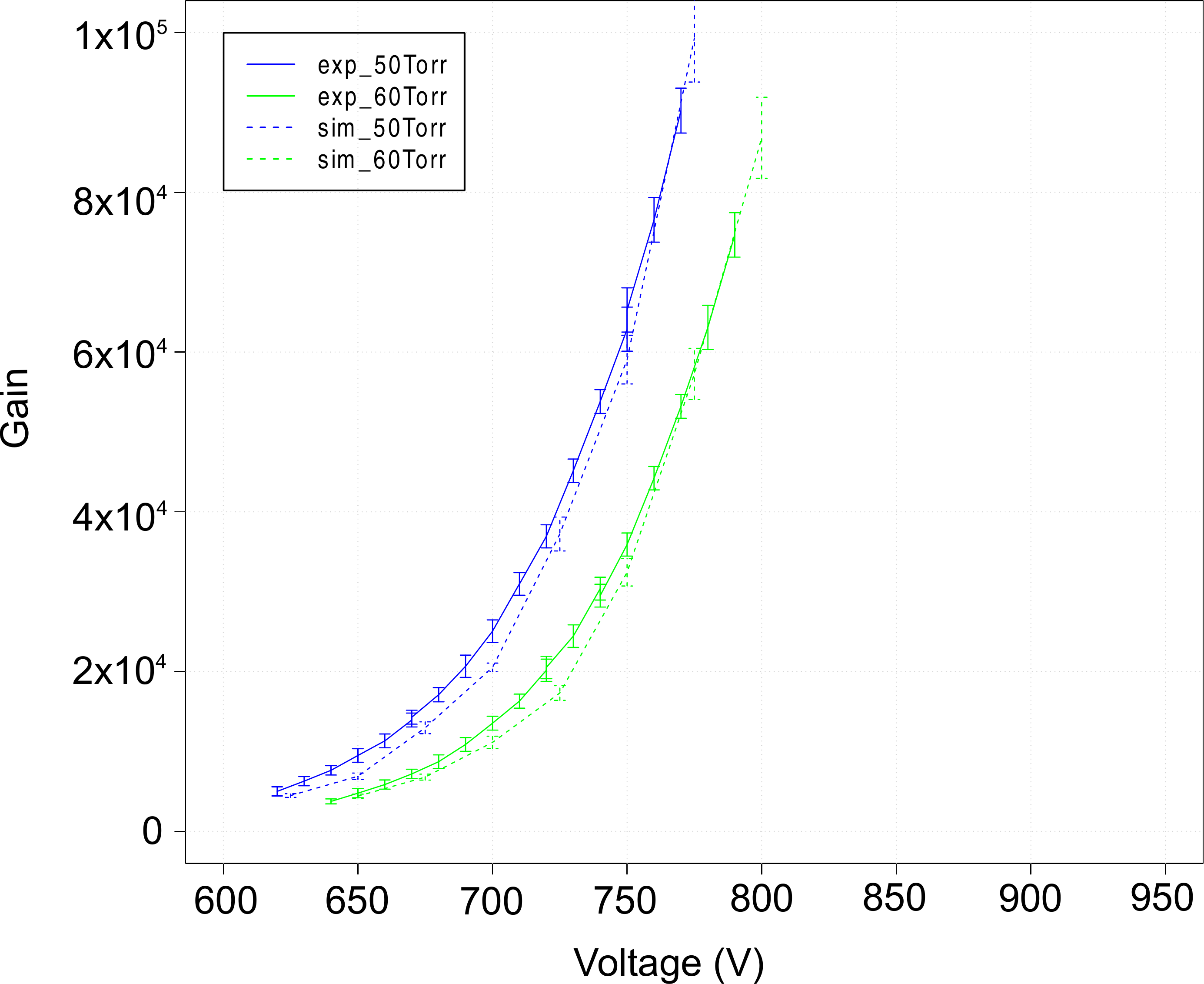}
\caption{Experimental (line) and simulated (dashed line) gain at 50 (blue) and 60~Torr (green) for the CERN THGEM}
\label{fig:Cern_gain}
\end{figure}

Garfield++ \cite{Baraka2015} was used to simulate the gain for comparison with measured results; this was first done for the well-characterised CERN THGEM and compared against the CF$_4$ gain measurements taken using this THGEM to validate the simulations. Figure \ref{fig:Cern_gain} shows a comparison between experimental and simulated data for the CERN THGEM at 50 and 60~Torr CF$_4$. It can be seen that the simulation is in close agreement with the experimental data.
 
\begin{figure}[htbp]
\centering 
\includegraphics[height=5cm]{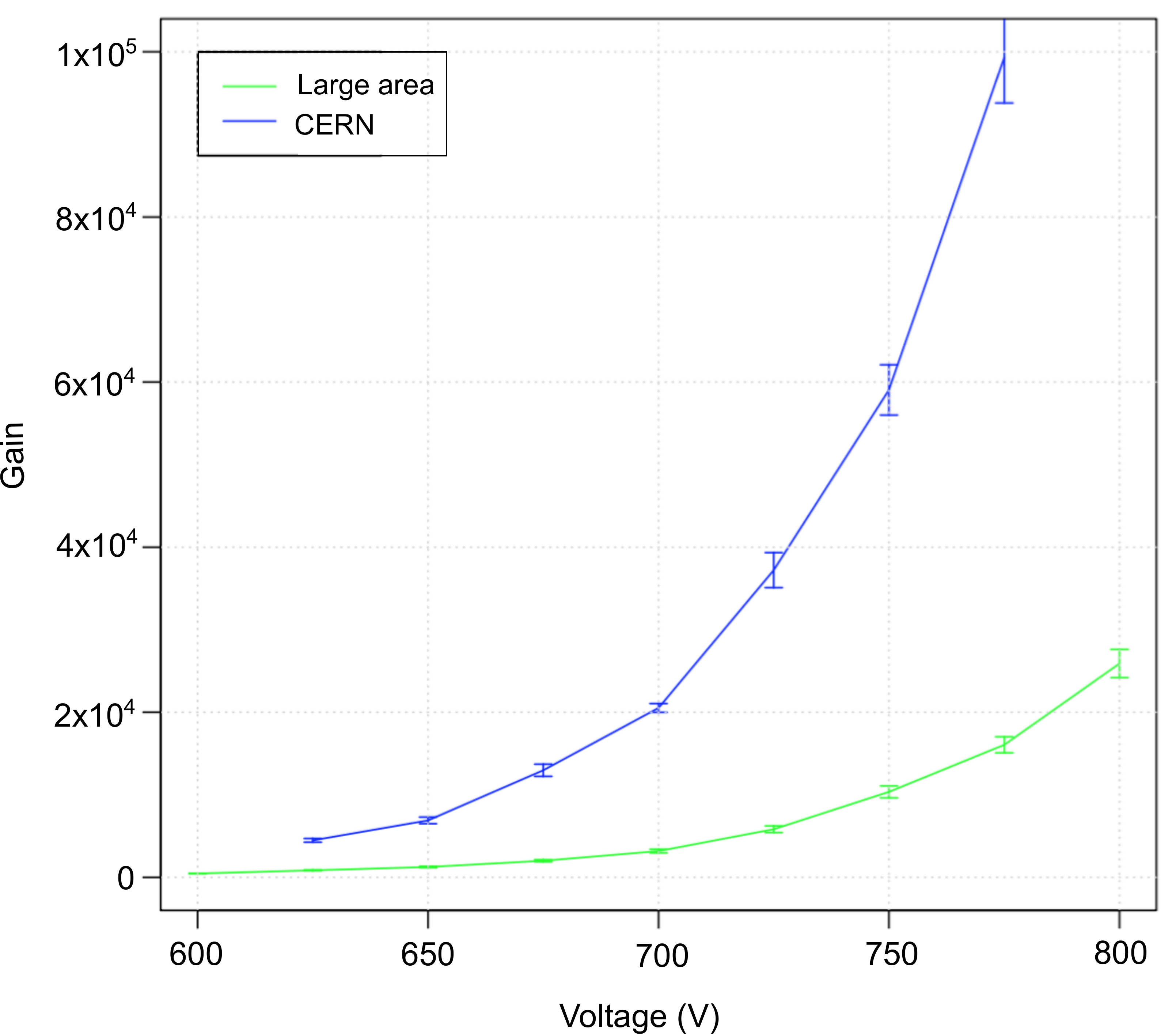}
\includegraphics[height=5cm]{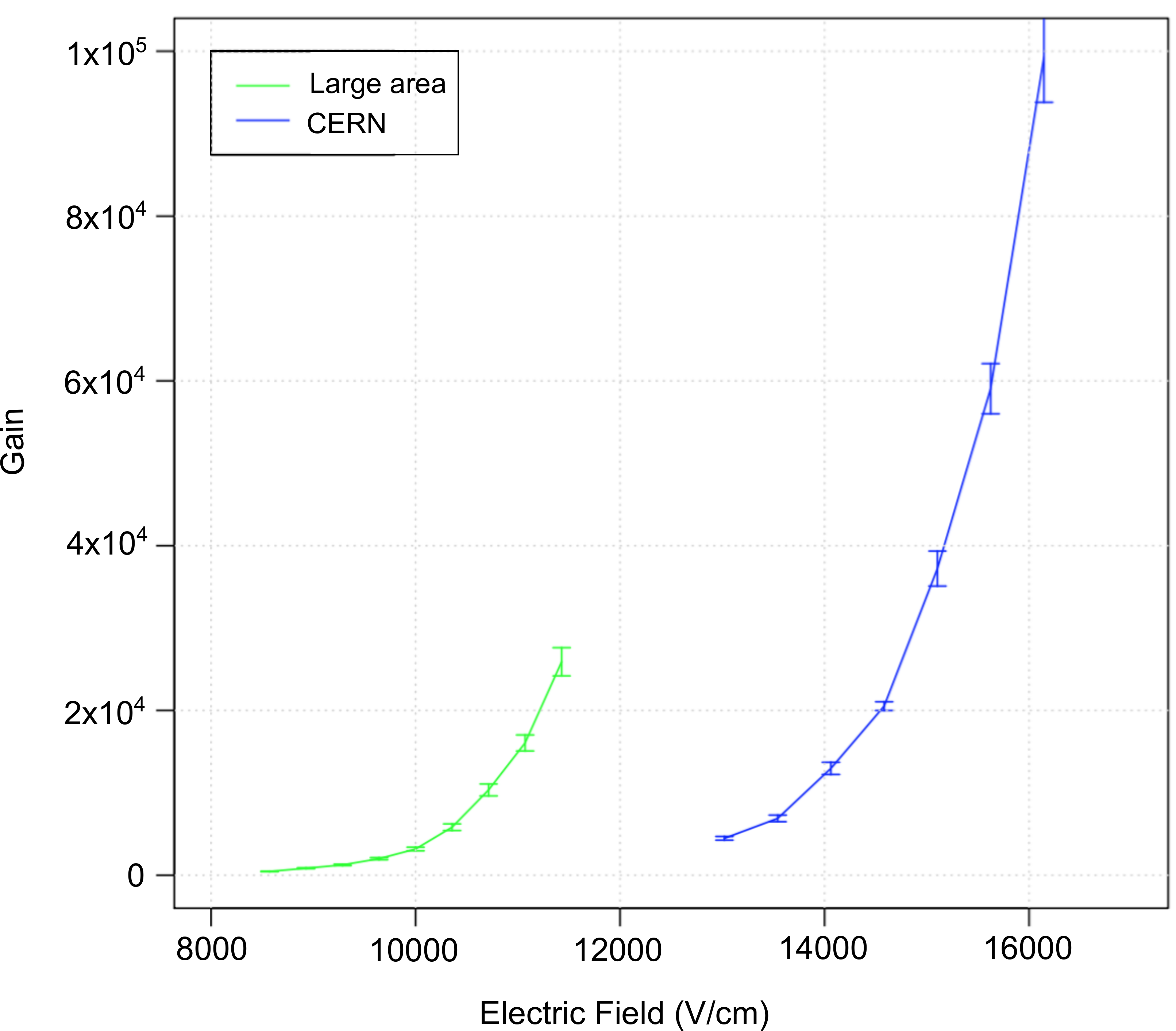}
\caption{Simulated gain shown as a function of voltage (left) and electric field (right) for the CERN and large area THGEMs at 50~Torr CF$_4$}
\label{fig:gain_v_e}
\end{figure}

The simulation was then run for the larger THGEM. Figure \ref{fig:gain_v_e} (left) shows a gain comparison between the CERN and 50~cm$^2$ THGEMs at 50~Torr plotted against voltage and Figure \ref{fig:gain_v_e} (right) shows the same result plotted against electric field.
 
Figure \ref{fig:gain_v_e} (left) shows a smaller gain for the large area THGEM when compared to the gain at the same voltages as the CERN THGEM. However, the THGEM gain is higher for this simulation than was measured from
the alpha spectra. For example the simulation at 700~V suggests the gain should be in the thousands rather than in the hundreds as derived from measurement. This may be due to imperfect collimation or the pre-amplifier becoming saturated.

The simulations also suggest that the large area THGEM should have a higher gain for weaker electric fields than the CERN THGEM (see Figure \ref{fig:gain_v_e} (right)). This may be due to the pitch of the large THGEM being twice that of the CERN THGEM and therefore having a greater metal area to collect avalanched electrons. 

\section{Future Detector Development}

There are several streams of work towards the creation of a CYGNUS prototype directional dark matter detector \cite{Spooner2017}. One element of the research and development is the construction of a back-to-back THGEM detector at Boulby Underground Laboratory, whilst a multi-readout detector for comparison of readout technologies is being created at Kobe University in Japan \cite{Miuchi2017}.

\subsection{Boulby detector}

\begin{figure}[htbp]
\centering
\includegraphics[height=60mm]{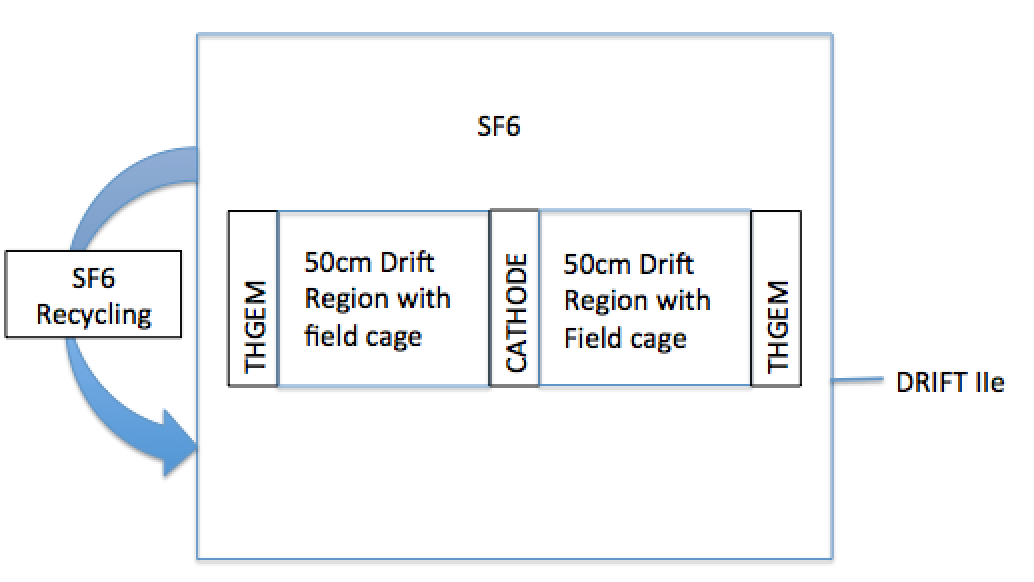}
\caption{Simplified schematic showing a possible future CYGNUS THGEM prototype at Boulby}
\label{fig:Boulby_detector}
\end{figure}

The proposed detector at Boulby, shown in Figure \ref{fig:Boulby_detector}, would consist of two back-to-back large area THGEMs similar to those used in this work, a central cathode and two 50~cm drift regions. Two field cages, with 50~cm drift distances, run from the central cathode towards the THGEMs and will step the voltage down steadily in order to maintain a uniform drift field. A uniform drift field ensures that accurate positional information is retained for an ionisation event. Figure \ref{fig:Boulby_detector} is an illustration of the basic setup for this type of detector.

\begin{figure}[htbp]
\centering
\includegraphics[height=50mm]{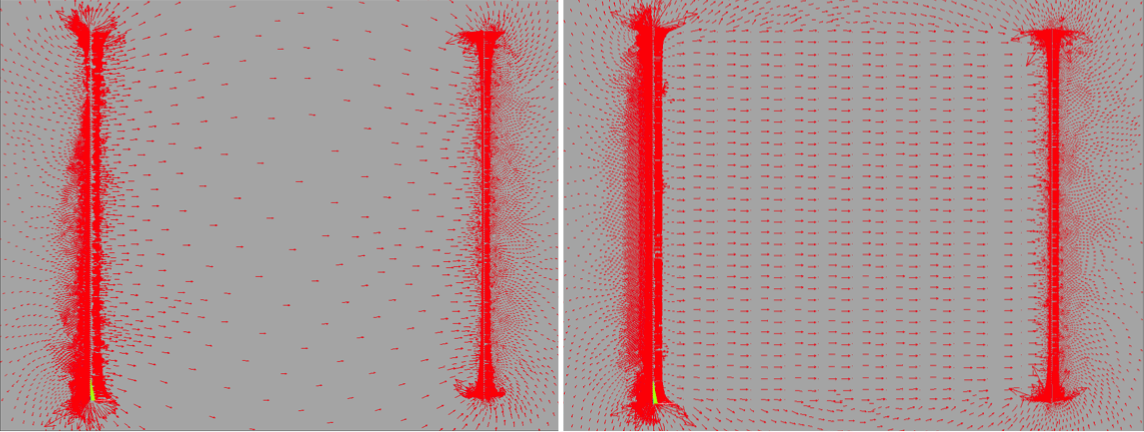}
\caption{Cross-sectional electric field distribution within the field cage for the 50~cm drift region of the full-scale design before (left) and after (right) the fieldcage is implemented.}
\label{fig:with_without_cage}
\end{figure}

A basic Ansys \cite{2012} simulation shown in Figure \ref{fig:with_without_cage} shows the drift field with and without a field cage. The field cage simulated in this Figure is over 50~cm in the drift direction and is made from 11 copper rings. The field demonstrates a high level of uniformity across the device, with increasing field strengths closer to the field rings.

A 9~cm prototype field cage has been created in the workshop at the University of Sheffield and will soon be tested with the 50~cm~by~50~cm THGEM. The prototype field cage is pictured in Figure \ref{fig:fieldcage} below. 

\begin{figure}[htbp]
\centering
\includegraphics[height=60mm]{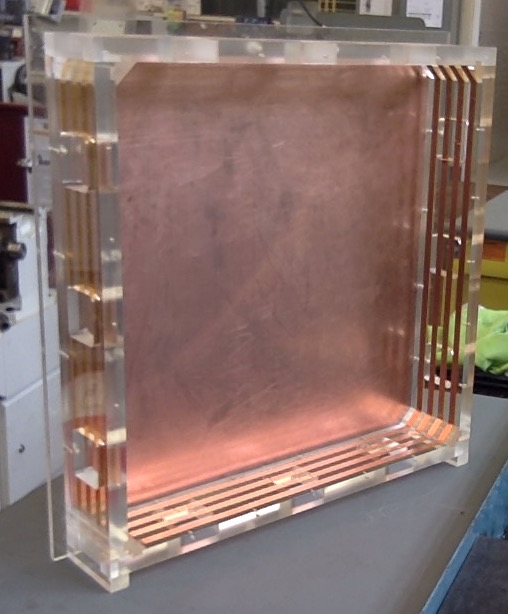}
\caption{Prototype field cage with a 9~cm drift length, soon to be tested with large area THGEMs.}
\label{fig:fieldcage}
\end{figure}

\subsection{CYGNUS-KM detector}

\begin{figure}[htbp]
\centering
\includegraphics[height=60mm]{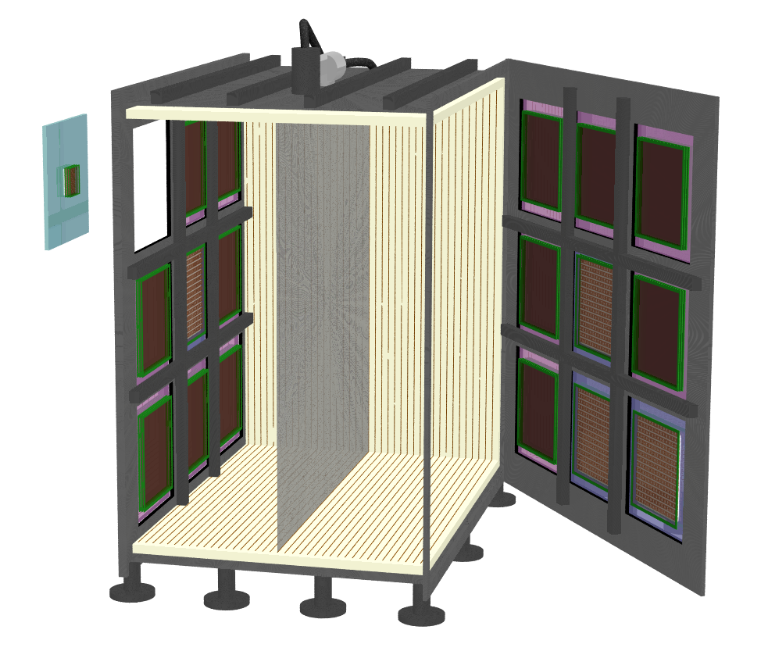}
\caption{Illustration of the CYGNUS-KM vessel. Nine readout planes are tiled to cover the vessel's cross-section.}
\label{fig:CYGNUS-KM}
\end{figure}

The CYGNUS-KM detector, shown in Figure \ref{fig:CYGNUS-KM}, is currently under construction at Kobe University, Japan. There are spaces for nine different detector readouts on either side of the cathode. CYGNUS-KM will test a variety of readouts within an SF$_6$ gas environment to see which would be most suited for scaling up to the large TPC volume of CYGNUS. The vessel will compare such properties as electron/nuclear recoil discrimination, fiducialisation, track direction reconstruction and gain. The University of Sheffield has agreed to supply at least one of the detector readouts for comparison which will most likely be a THGEM of similar dimensions to those describe in this work.

\section{Conclusions and further work}

We have reported experimental measurements of the Townsend coefficients using a THGEM and the operational demonstration of 50~cm~by~50~cm THGEMs.

We have described an optical inspection system used to identify fabrication imperfections on THGEMs which may cause sparks and hence limit the operational gain of the detector. The technique has been used to quantify manufacturing tolerances to aid design
of imperfection-tolerant THGEM architectures.

Experimentally measured gains are significantly lower than those predicted by computational simulations. Further work will seek to understand these discrepancies and optimise the large area THGEMs.

\appendix

\acknowledgments

The authors wish to thank Jez Boakes and Trevor Gamble for engineering support and Chris Toth for his assistance at Boulby Underground Laboratory. Thanks also go to Ross Hunter for his early contribution to this work. We thank Dinesh Loomba and Nguyen Phan for helpful discussions on THGEM operation. We acknowledge support from the STFC through grant number ST/P00573X/1, the Ministry of Defence and the Home Office OSCT.


\bibliographystyle{JHEP}
%% Enter the path to your bibliography
\bibliography{Library}

\end{document}